# Fabry-Pérot Cavity formed by Magnonic waveguide using holes and notches

Rajesh Kumar[1], Yat-Yin Au[1], and Snorri Ingvarsson[1]

[1]Science Institute, University of Iceland, Dunhaga 3, IS-107 Reykjavik, Iceland

**In this paper, we have investigated a spin-wave-based Fabry-Pérot (FP) cavity using two air holes and notches embedded in a magnonic waveguide. Through micromagnetic simulations, we have demonstrated the spin-wave profile, as well as the transmission and reflection spectra of the FP type magnonic waveguide, by varying the distance between the center of holes and notches. We have introduced an analysis of the reflection spectra in an FP type magnonic waveguide, which is essential and considerable for the performance of magnonic integrated circuits. We have also presented a theoretical model to describe the transmission and reflection characteristics of the FP-type magnonic waveguide. Furthermore, active tuning of the resonance peak of the FP type magnonic wavguide by varying the external magnetic field has been demonstrated. Our proposed structure and its study may contribute to the development of magnonic integrated circuits for low-power computing and data processing in next-generation spintronic technologies.**

## I. Introduction

A Fabry-Pérot cavity (FPc) consists of two parallel reflecting surfaces, where waves transmit through the cavity when the incoming waves constructively interfere with the waves trapped inside it, while destructive interference results in no transmission[1]. In optics, FPc are widely used in telecommunication networks, filters, and sensors. Compared to electromagnetic waves, the wavelength of spin waves (the dynamic excitations of spins in magnetic materials) are in the nanoscale at GHz frequencies, enabling device miniaturization down to nanometers [2][3]ultralow losses, and the exploration of abundant nonlinear phenomena [4]. These features make spin waves promising candidates as data carriers for future low-energy data processing [5] and neuromorphic computing [6], [7].

A magnonic FPc, analogous to an optical FPc, provides an efficient way to control spin-wave propagation on the nanoscale at GHz frequencies[8] with a simple and compact structure. A magnonic FPc can be realized by creating two parallel reflecting interfaces in a low-loss magnetic material, which enables long decay times and extended spin-wave propagation. Among low-loss magnetic materials, yttrium iron garnet (YIG) is a promising candidate due to its low damping factor, which supports long-range spin-wave propagation [9]. Parallel reflecting surfaces in a magnonic waveguide (made of YIG) can be formed by placing patterned ferromagnetic materials or by using the Oersted field of a current-carrying wire on top of the magnonic waveguide[10]. However, these designs have complex structures, require complicated fabrication processes, and offer limited fabrication tolerance. Secondly, due to the presence of interfacing surfaces, spin-wave propagation occurs in both directions, and reverse propagation becomes significant; therefore, reflection analysis becomes essential for the device. In previous literature, only forward propagation of the spin wave in FP type magnonic waveguide was discussed[11], [12], [13], [14], [15], while backward propagation i.e reflection was ignored.

Here, we have introduced two magnonic FP nanocavities: the first with two air holes and the second with two notches (both act as reflecting mirrors) in a magnonic waveguide, enabling flexible control of spin-wave propagation. In this paper, we have analyzed the transmission and reflection spectra by varying the distance between the two reflecting surfaces created by the holes and notches inside the magnonic waveguide. A theoretical model has been developed to describe the transmission and reflection of the waveguide. Furthermore, we have examined the transmission and reflection spectra by varying the external magnetic field. Thus, by tuning the external magnetic field (active tuning), spin-wave propagation has been controlled, which is essential for beyond-CMOS technology, data processing and magnonic computing.

## II. Design and material of FP Type magnonic waveguide

We have designed two FPc types on a magnonic waveguide: the first incorporating two air holes (FP-H) and the second incorporating two notches (FP-N). We have included a reference waveguide (without air holes) with two ports defined, the input port positioned on the left and the output port positioned on the right side of the waveguide (Fig. 1a). Two air holes were introduced on either side of the waveguide center, maintaining equal separation from the central region to ensure structural symmetry (Fig. 1b). The distance between the two air holes or notches is denoted as $d$, and the radius of each hole and semi-hole is set to 20 nm. The notches were formed by removing semicircular regions (radius $R$) from the two opposing sides of the waveguide at two different positions with separation $d$, as shown in Fig. 1(c), positioned symmetrically about the center of the magnonic waveguide at equal distances. In both the FP-H and FP-N structures, the two air holes or notches are positioned symmetrically about the center of the waveguide, each located at a distance of $d/2$ from the center (Fig. 1b and 1c). The width of the waveguide is $w = 100$nm, and the lengths of both magnonic waveguides are identical, with $L = 2200$nm. The heights of both the waveguides, air holes and notches are the same, with a value of $h = 30$nm. Yttrium iron garnet (YIG) has been chosen as the material for the waveguides. The magnetization parameters of YIG are taken from experimental data, such as the exchange constant $A_{ex} = 3.5\, pJm^{-1}$, saturation magnetization $M_s = 1.4 \times 10^5\, A/m^{-1}$ and the Gilbert damping factor of the waveguide is $\alpha = 2 \times 10^{-4}$[16].

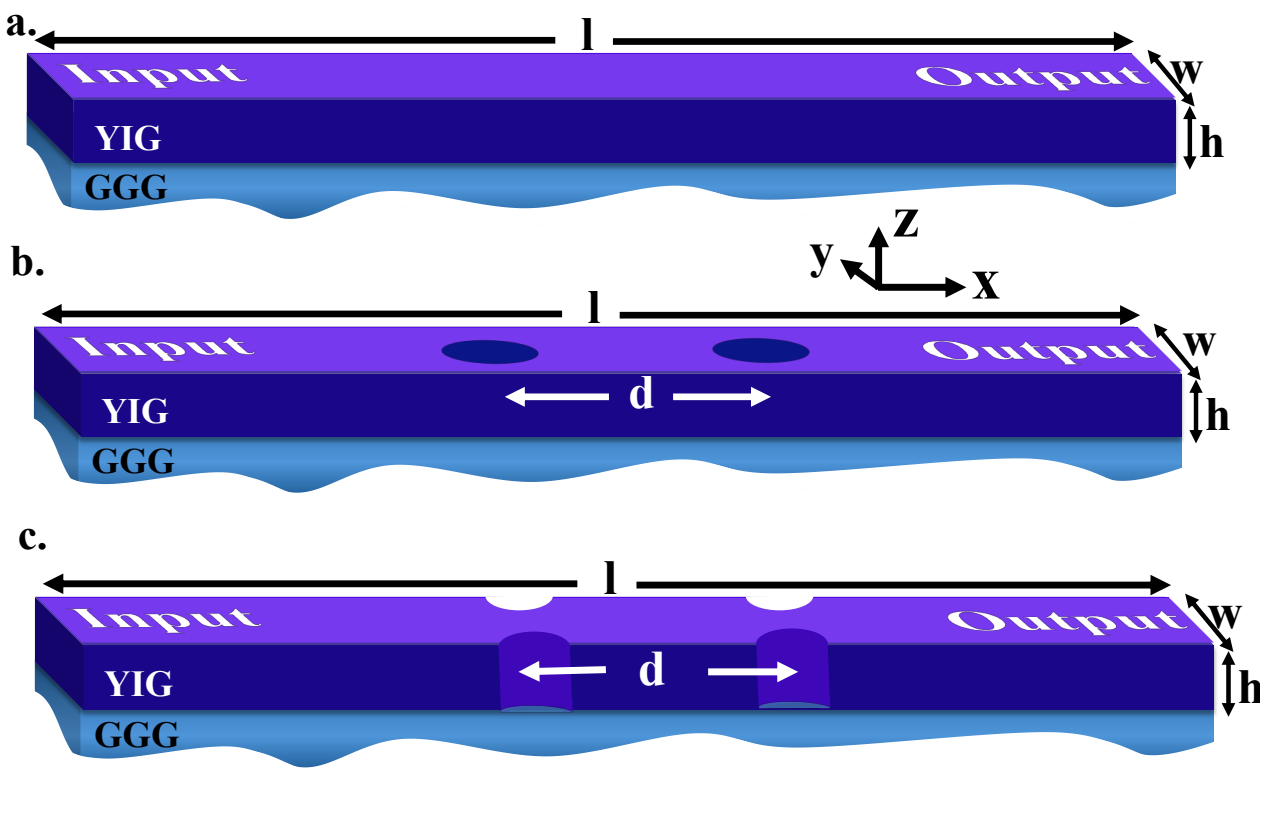

**Figure 1.** Perspective views of the FP type magnonic waveguides showing the ports, structural parameters, and materials for: (a) the reference waveguide, (b) the FP waveguide with holes (FP-H), and (c) the FP waveguide with notches (FP-N).

## III. Theoretical Discussion

The theoretical derivation of the transmission and reflection of a FP-type magnonic waveguide can be described using the standard method applied to photonic [17], [18]. The two air holes (of the same radius) or two notches (of identical dimensions) act as partially reflecting planes for the propagating spin waves in the magnonic waveguide. Thus, the FP type magnonic waveguide supports both forward (backward) propagating modes at the input and output, with complex spin-wave amplitudes $A_{F_{in}}(A_{B_in})$ and $A_{F_{out}}(A_{B_out})$ as shown in Fig. 2a. When the spin waves propagate through the reflecting plane (plane1 (plane2)), partial reflection occurs with a reflection coefficient of r ($r_1(r_2)$) phase jump of $\theta$ ($\theta_1(\theta_2)$). The transfer matrix equation relating the incoming and outgoing spin-wave amplitudes for the FP type magnonic waveguide can be written as:

$$\begin{bmatrix} A_{F_out} \\ A_{B_in} \end{bmatrix} = \frac{i}{\sqrt{1-{r_2}^2}} \begin{bmatrix} 1 & r_2 \\ -r_2 & -1 \end{bmatrix} \begin{bmatrix} e^{-i\theta_2} & 0 \\ 0 & e^{i\theta_2} \end{bmatrix} \begin{bmatrix} e^{-i\theta_1} & 0 \\ 0 & e^{i\theta_1} \end{bmatrix} \frac{i}{\sqrt{1-{r_1}^2}} \begin{bmatrix} 1 & r_1 \\ -r_1 & -1 \end{bmatrix} \begin{bmatrix} A_{F_in} \\ A_{B_out} \end{bmatrix} \quad ......(1)$$

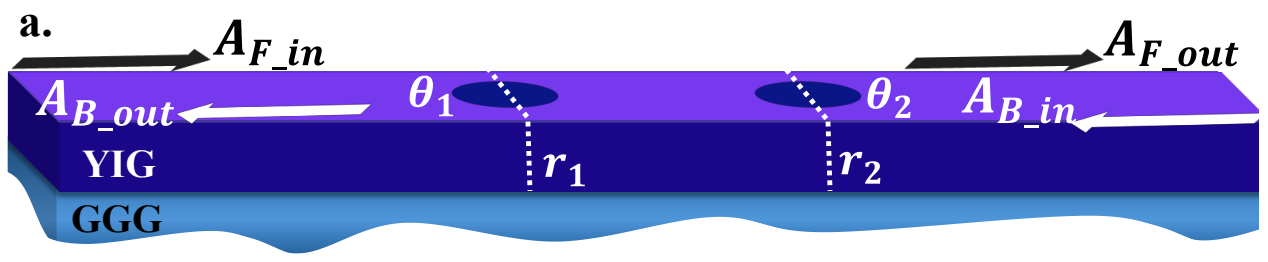

**Figure 2.** (a) Schematic view illustrating the phase accumulation, reflection coefficients due to the holes, and the input and output spin-wave amplitudes at both ends of the waveguide.

To simplify, only the incoming spin-wave amplitude from the left input is considered, so the input from the right is set to zero ($A_{B_in} = 0$). The expressions for the amplitude transmission ($T$) and reflection ($R$) can then be calculated as:

$$T = \frac{A_{F_out}}{A_{F_in}} = \frac{\sqrt{1-{r_1}^2}\sqrt{1-{r_2}^2}\, e^{i\theta}}{r_1 r_2 - e^{2i\theta}} \quad ...... (2)$$

$$R = \frac{A_{B_out}}{A_{F_in}} = \frac{-r_2 + r_1 e^{2i\theta}}{r_1 r_2 - e^{2i\theta}} \quad ...... (3)$$

In figure 3, we have presented the theoretical transmission and reflection spectra of the FP type magnonic waveguide. For this purpose, the spin-wave dispersion relation was evaluated for a 100 nm-wide magnonic waveguide, providing the relationship between wavenumber and frequency. The phase accumulation was calculated using $\theta = d \times k$, where $k$ is the spin-wave wavenumber. By substituting the values of $r_1$, $r_2$, and $\theta$, in the equation (2) and equation (3), the theoretical transmission and reflection spectra were obtained, as shown in Fig. 3(a) and Fig. 3(b), respectively.

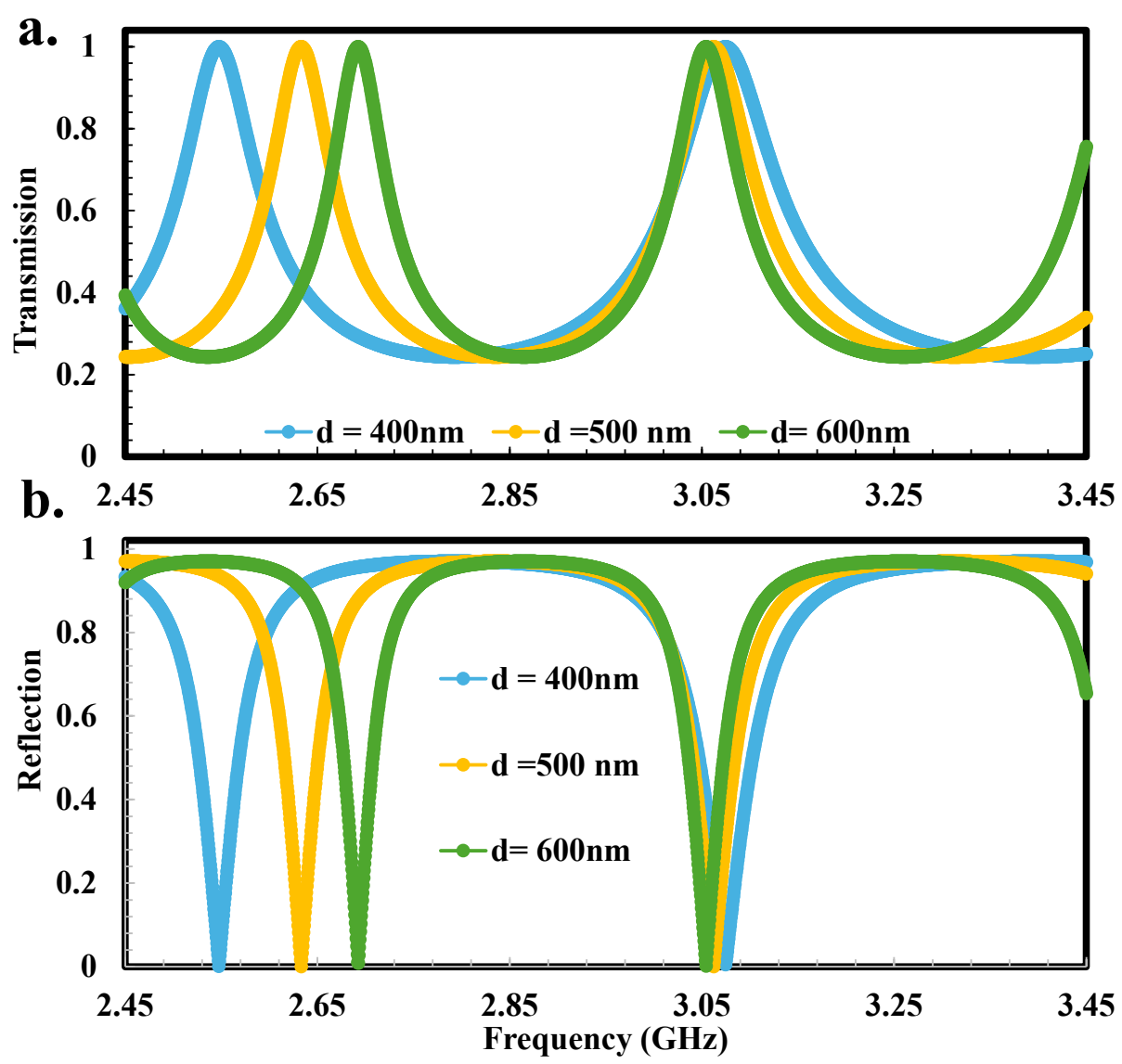

Figure 3. (a) Theoretical transmission spectrum of the FP typed magnonic waveguide. (b) Corresponding reflection spectrum. The reflection coefficients were taken as $r_1 = r_2 = 0.81$.

## IV. Micromagnetic Simulation for FP-H and FP-N

For the investigation of FP-H and FP-N, micromagnetic simulations were performed using the open-source software MuMax3, which employs the finite-difference method to solve the Landau–Lifshitz–Gilbert equation [19]. Two different approaches were applied for the analysis of transmission and reflection: first, broadband pulse excitation (BPE) for obtaining the transmission and reflection spectra, and second, continuous wave excitation (CWE) for analyzing the spin-wave profile at a single frequency. The average cell size was set to $10 \times 10 \times 10\, nm^3$ for both approaches. The initial magnetization of the waveguide was oriented along the x-axis, due to strong shape anisotropy resulting from the small width relative to its length, and the system was allowed to relax to obtain the ground state.

The calculation of transmission and reflection of the structure was carried out in three parts.

In the first two parts, a reference waveguide (REW) and a FP type waveguide (FPW) were considered. In both waveguides, an increasing damping factor was applied at the opposite end, from $\alpha$to $\alpha_{\mathrm{damp}}$, to eliminate spin-wave reflections at the ends (Fig. 4a and 4b). To calculate the reflection of the FP type waveguide, a perfect reflecting waveguide (PRW) was introduced in the third part, in which the damping factor ($\alpha$) is the same as in the first two parts for the left half, while for the right half, $\alpha$is increased to 100 (Fig. 4c). The excitation region ($x_E$) and the input ($x_1$) and output ($x_2$) chosen points are indicated in figure 4, with their values provided in Table I. To generate spin waves in the waveguide, a sinc pulse was used for broadband pulse excitation, as described in Table I.

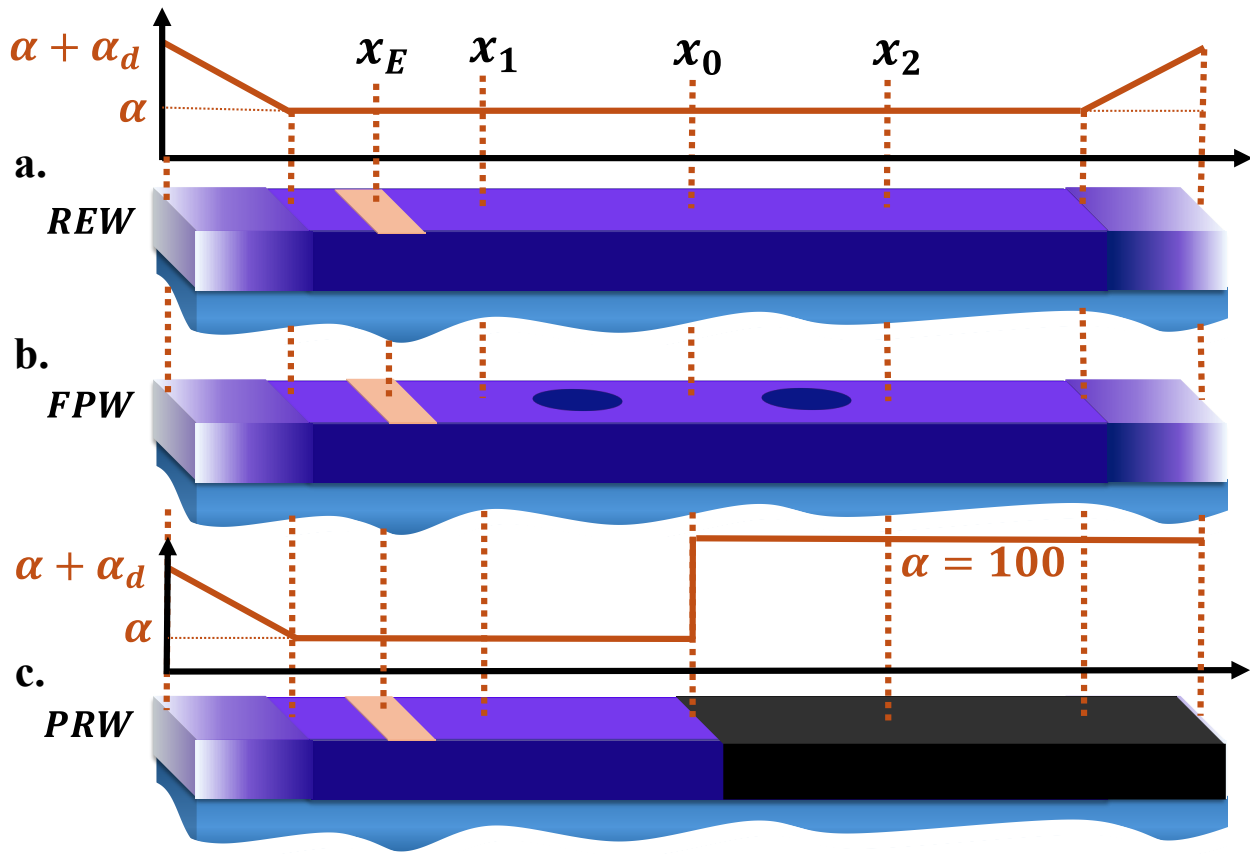

**Figure 4.** Three-dimensional views of the micromagnetic simulation structures: (a) reference waveguide (REW), (b) FP type waveguide (FPW), and (c) perfect reflecting waveguide (PRW). The orange strip on the left side of each waveguide represents the excitation region. $\alpha_d$and $x_0$indicate the increased damping values applied at the left and right sides of the REW and FPW, and only at the left side of the PRW. The black region on the right side of the PRW represents a high-damping material ($\alpha = 100$).

TABLE I

PARAMETER OF MICROMAGNETIC SIMULATION

| Parameter | BPE | CWE |
|---|---|---|
| $x_{total}$ | $5\ \mu m$ | 2200 nm |
| $x_0$ | $2.5\ \mu m$ | 1100 nm |
| $x_1$ | $2.1\ \mu m$ | 700 nm |
| $x_2$ | $2.9\ \mu m$ | 1500 nm |
| $x_E$ | $1.2\ \mu m$ | 500 nm |
| $h_y(x,t)$ | $h_0 \dfrac{\sin(\omega_c(t-t_0))}{\omega_c(t-t_0)} \omega \dfrac{\sin(k_c(x-x_E))}{k_c(x-x_E)}$ | $h_0 \sin(\omega t)$ |

The simulation was run for 100 ns, and the $z$-component of the magnetization was recorded, denoted as: $m_{z,FPW}(x_1,t)$, $m_{z,FPW}(x_2,t)$, $m_{z,REF}(x_1,t)$, $m_{z,REF}(x_2,t)$ $m_{z,PRW}(x_1,t)$, and $m_{z,PRW}(x_2,t)$. For the BPE mode, the transmission ($T_{\mathrm{pulse}}$) and reflection ($R_{\mathrm{pulse}}$) are calculated as:

$$T_{pulse} = \frac{F(m_{z,FPW}(x_2,t))(f)}{F(m_{z,REW}(x_2,t))(f)}$$

$$R_{pulse} = -\frac{F(m_{z,FPW}(x_1,t) - m_{z,REW}(x_1,t))(f)}{F(m_{z,PRW}(x_1,t) - m_{z,REW}(x_1,t))(f)}$$

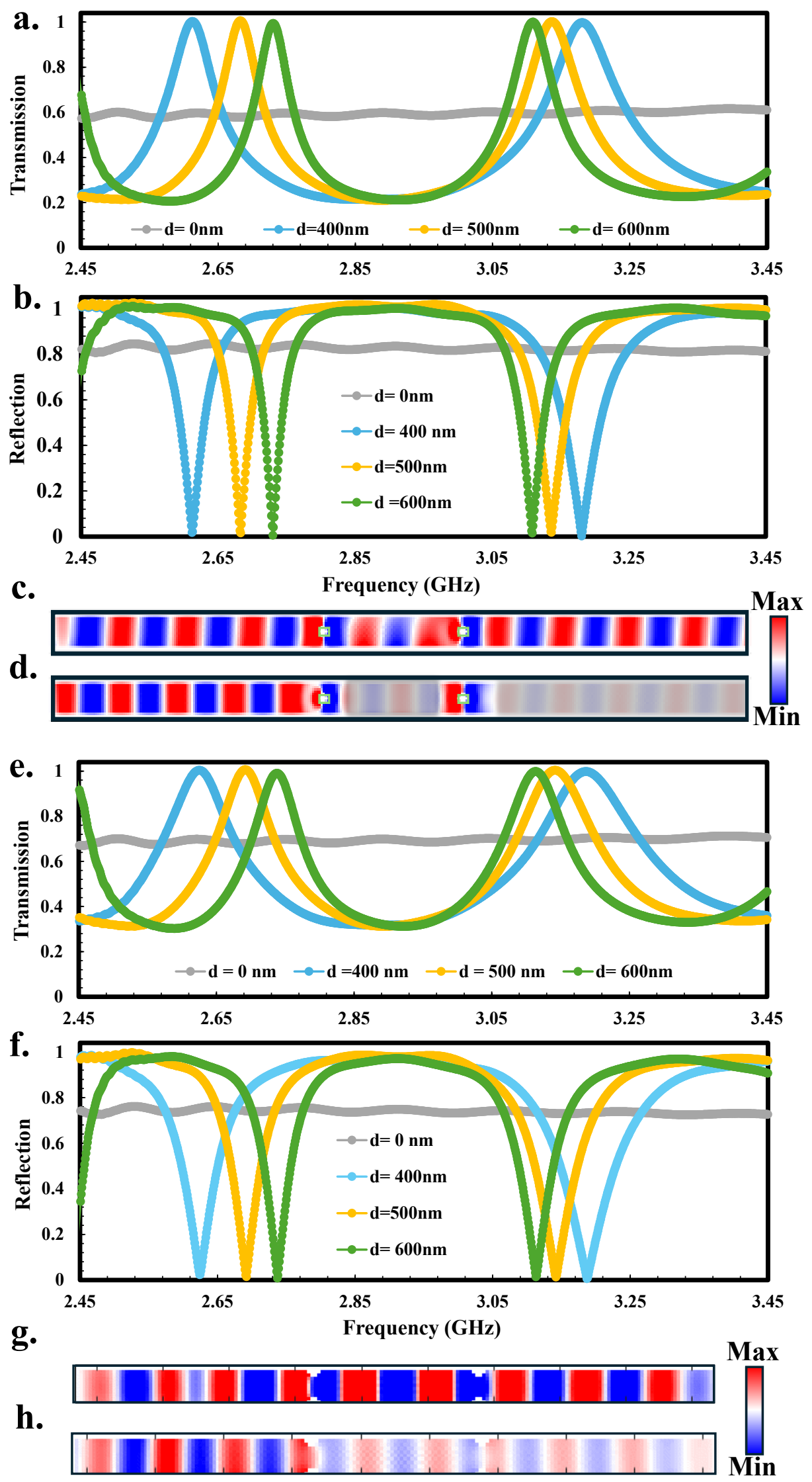

**Figure 5.** Transmission and reflection characteristics with varying cavity lengths ($d$). (a) Transmission spectra of FP-H for different values of $d$. (b) Reflection spectra of FP-H for different values of $d$. Spin-wave profiles of FP-H at $d = 500$nm: (c) at the resonance peak ($f = 2.686$ GHz) and (d) at the off-resonance frequency ($f = 2.928$ GHz). (e) and (f) Transmission and reflection spectra of FP-N for different values of $d$. Spin-wave profiles of FP-N at $d = 600$nm: (g) at the resonance peak and (h) at the off-resonance frequency.

Where $F$ is the Fourier transform. In case of CWE mode, the transmission ($T_{cw}$) and reflection ($R_{cw}$) are given as

$$T_{cw} = \frac{mag(m_{z,FPW}(x_2,t))}{mag(m_{z,REW}(x_2,t))}$$

$$R_{cw} = \frac{mag(m_{z,FPW}(x_1,t) - m_{z,REW}(x_1,t))}{mag(m_{z,PRW}(x_1,t) - m_{z,REW}(x_1,t))}$$

Where $mag$ is the amplitude of the temporal sinusoidal fit of $m_z(t)$ when the system reached to the steady state.

## V. Result and discussion

### *A. Transmission Spectra*

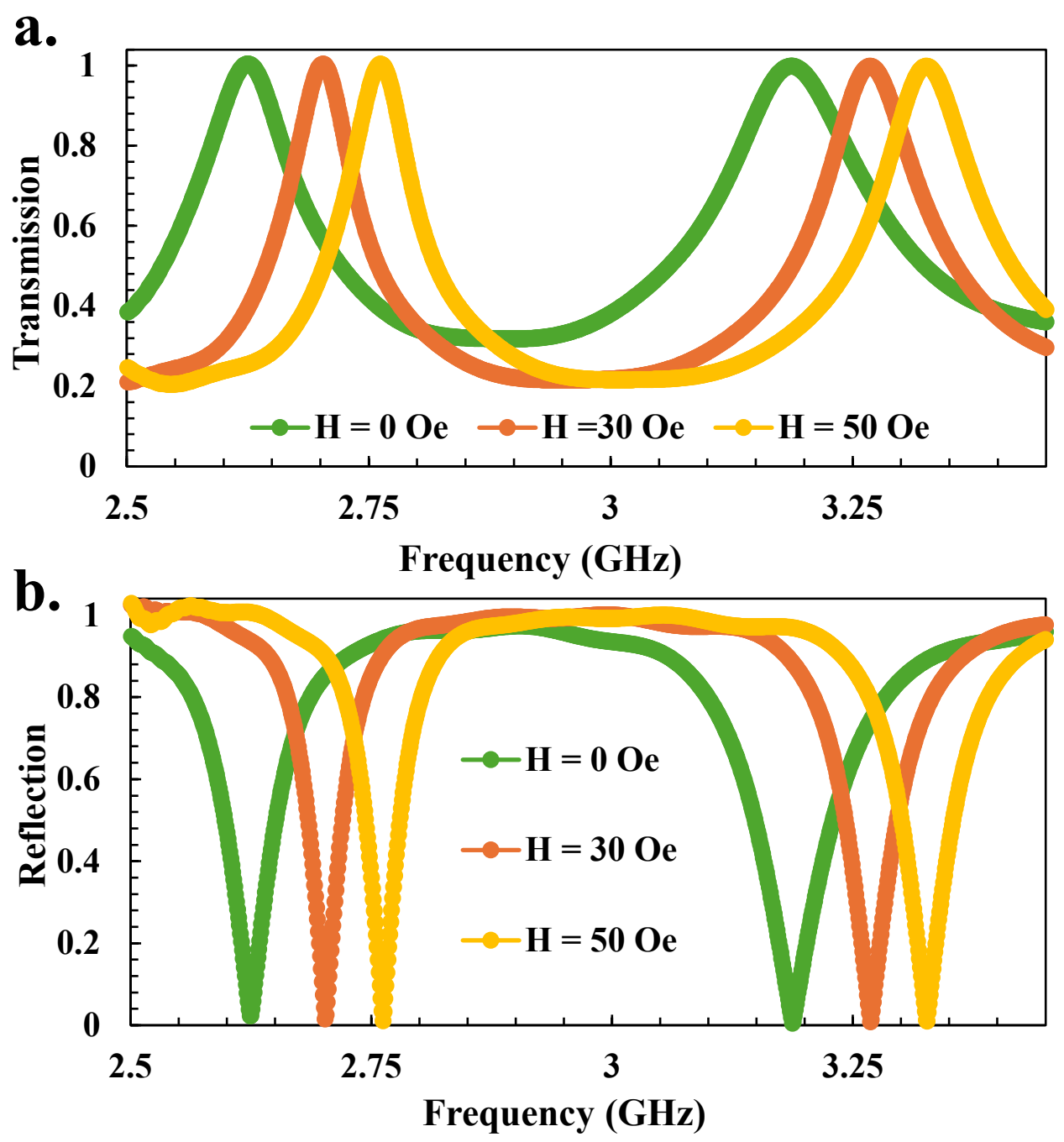

**Figure 6.** Transmission and reflection spectra of the FP-H under different external magnetic fields ($H$). (a) Transmission, (b) Reflection.

For the transmission and reflection spectra of the FP-H and FP-N with different values of $d$(0 nm, 400 nm, 500 nm, and 600 nm), a frequency sweep was performed from 2.45 GHz to 3.45 GHz, and multiple resonance peaks were observed for all values of $d$except $d$ = 0nm (Fig. 5a and 5b, grey curve), since at $d$ = 0nm, the hole and notch behave as point reflectors rather than as a cavity. Increasing $d$results in a decrease in the separation between the resonance peaks, and consequently, the number of peaks within the given frequency range increases. For d=500nm, at the resonance peak ($f_{\text{res}}$ = 2.686 GHz), the transmission and reflection values at the output port are 0.9999 and 0.0001, respectively. This shows that the spin wave generated at the input constructively interferes with the circulating spin wave in the cavity, propagates toward the output, and results in maximum transmission, as observed in the spin-wave profile shown in Fig. 5(c). However, at the off-resonance frequency ($f$ = 2.928 GHz), the transmission and reflection values are 0.21 and 0.99, respectively, indicating that the generated spin wave undergoes destructive interference, resulting in minimal transmission, as shown in Fig. 5(d). The behavior of the spin-wave profile, transmission, and reflection spectra for the FP-N is similar to that of the FP-H, as shown in Figs. 5.

### *B. Tuning of resonance peak by external magnetic field*

To analyze the transmission and reflection spectra of FP-H for different values of the external magnetic field ($H$), a frequency sweep was performed from 2.45 GHz to 3.45 GHz at external fields of 0 mT, 30 mT, and 50 mT, with $d$ fixed at 500 nm. Increasing the external magnetic field results in a shift of the resonance peak (dip) in the transmission spectra (reflection spectra), as shown in Fig. 6(a) and 6(b) respectively. An increase in the external magnetic field leads to an increase in the spin-wave frequency at a constant wavenumber. Therefore, for a fixed cavity length, the adjacent resonance peaks or dips in the transmission and reflection spectra, respectively, shift toward higher frequencies. Similar results have been obatined for the FP-N (not shown here).

## VI. Conclusion

In conclusion, the transmission and reflection spectra for FP-H and FP-N at different separations $d$ and the spin-wave profiles have been analyzed using micromagnetic simulations. For the first time, we have introduced reflection analysis for the FP type magnonic waveguide. Active tuning of the resonance peak of the FP-H by varying the external magnetic field has been demonstrated, which may support magnonic computing and beyond-CMOS technologies. Our proposed structure not only operates on the nanoscale at GHz frequencies but is also easy to fabricate and offers strong fabrication tolerance, which may help in building compact magnonic integrated circuits for data-processing systems and computing.

## Acknowledgement

This work is supported by funding from the Icelandic Research Fund Grant No. 228951 and 2410333.

**Disclosures.** The authors declare no conflicts of interest.

## References

[1] G. Hernández, *Fabry-perot interferometers*, no. 3. Cambridge University Press, 1988.

[2] A. V Chumak, V. I. Vasyuchka, A. A. Serga, and B. Hillebrands, “Magnon spintronics,” *Nat. Phys.*, vol. 11, no. 6, pp. 453–461, 2015.

[3] S. Neusser and D. Grundler, “Magnonics: Spin waves on the nanoscale,” *Advanced materials*, vol. 21, no. 28, pp. 2927–2932, 2009.

[4] R. Dreyer, A. F. Schäffer, H. G. Bauer, N. Liebing, J. Berakdar, and G. Woltersdorf, “Imaging and phase-locking of non-linear spin waves,” *Nat. Commun.*, vol. 13, no. 1, p. 4939, 2022.

[5] Y.-Y. Au and K. G. Fripp, “Electric field control of chiral magnonic resonators for spin-wave manipulation,” *Phys. Rev. Appl.*, vol. 20, no. 3, p. 034023, 2023.

[6] Y.-Y. Au, “Electric chiral magnonic resonators utilizing spin–orbit torques,” *J. Appl. Phys.*, vol. 135, no. 4, 2024.

[7] Á. Papp, W. Porod, and G. Csaba, “Nanoscale neural network using non-linear spin-wave interference,” *Nat. Commun.*, vol. 12, no. 1, p. 6422, 2021.

[8] H. Qin *et al.*, “Nanoscale magnonic Fabry-Pérot resonator for low-loss spin-wave manipulation,” *Nat. Commun.*, vol. 12, no. 1, p. 2293, 2021.

[9] H. Yu *et al.*, “Magnetic thin-film insulator with ultra-low spin wave damping for coherent nanomagnonics,” *Sci. Rep.*, vol. 4, no. 1, p. 6848, 2014.

[10] M. P. Kostylev *et al.*, “Resonant and nonresonant scattering of dipole-dominated spin waves from a region of inhomogeneous magnetic field in a ferromagnetic film,” *Physical Review B—Condensed Matter and Materials Physics*, vol. 76, no. 18, p. 184419, 2007.

[11] I.-S. Hong, S. K. Kim, K.-J. Lee, and G. Go, “Tunable magnonic cavity analogous to Fabry–Pérot interferometer,” *Appl. Phys. Lett.*, vol. 119, no. 20, 2021.

[12] A. Lutsenko, K. G. Fripp, L. Flajšman, A. V Shytov, V. V Kruglyak, and S. van Dijken, “Magnonic Fabry–Pérot resonators as programmable phase shifters,” *Appl. Phys. Lett.*, vol. 126, no. 8, 2025.

[13] H. Qin, “Reconfigurable Magnonic Devices for Spin-Wave

Manipulation on the Nanoscale," in *2025 9th IEEE Electron Devices Technology & Manufacturing Conference (EDTM)*, IEEE, 2025, pp. 1–3.

[14] H. Qin *et al.*, "Nanoscale magnonic Fabry-Pérot resonator for low-loss spin-wave manipulation," *Nat. Commun.*, vol. 12, no. 1, p. 2293, 2021.

[15] M. S. Sarker *et al.*, "Sensitivity enhancement in magnetic sensor using CoFeB/Y3Fe5O12 resonator," *Sci. Rep.*, vol. 12, no. 1, p. 11105, 2022.

[16] C. Dubs *et al.*, "Low damping and microstructural perfection of sub-40nm-thin yttrium iron garnet films grown by liquid phase epitaxy," *Phys. Rev. Mater.*, vol. 4, no. 2, p. 024416, 2020.

[17] L. Gu *et al.*, "A compact structure for realizing Lorentzian, Fano, and electromagnetically induced transparency resonance lineshapes in a microring resonator," *Nanophotonics*, vol. 8, no. 5, pp. 841–848, 2019.

[18] J. Wang, J. Lin, P. Jin, S. Liu, and K. Zhou, "Fano resonance in a microring resonator with a micro-reflective unit," *Opt. Express*, vol. 31, no. 19, pp. 31587–31596, 2023.

[19] A. Vansteenkiste, J. Leliaert, M. Dvornik, M. Helsen, F. Garcia-Sanchez, and B. Van Waeyenberge, "The design and verification of MuMax3," *AIP Adv.*, vol. 4, no. 10, 2014.